\documentclass[12pt]{article}

\usepackage{amsmath,amsfonts}

\usepackage{slashed}

\usepackage[french,english]{babel}

\usepackage{graphicx}
\usepackage{color}

\usepackage[T1]{fontenc}

\usepackage{hyperref} 
\hypersetup{colorlinks=true, citecolor=blue, filecolor=black, linkcolor=blue, urlcolor=blue, pdfpagemode=UseNone}

\advance \textheight by \topskip
\makeatletter
\@addtoreset{equation}{section}
 
\makeatother

\newcommand{\be}{\begin{equation}}
\newcommand{\ee}{\end{equation}}
\newcommand{\bea}{\begin{eqnarray}}
\newcommand{\eea}{\end{eqnarray}}
\newcommand{\dd}{\partial}

\def\>{\rangle}
\def\<{\langle}

\begin{document}

\title{
{\bf On a classical solution to  the Abelian Higgs model }}


\author{
{\sf   N. Mohammedi} \thanks{e-mail:
noureddine.mohammedi@univ-tours.fr}$\,\,$${}$
\\
{\small ${}${\it Institut  Denis Poisson (CNRS - UMR 7013),}} \\
{\small {\it Universit\'e  de Tours,}}\\
{\small {\it Facult\'e des Sciences et Techniques,}}\\
{\small {\it Parc de Grandmont, F-37200 Tours, France.}}}
\date{}
\maketitle
\vskip-1.5cm

\vspace{2truecm}

\begin{abstract}

\noindent
A particular solution to the equations of motion of the Abelian Higgs model is given.
The solution involves the Jacobi elliptic functions as well as the Heun functions.

\end{abstract}

\newpage

%

\setcounter{equation}{0}

\section{Introduction}

The search for analytical solutions to the classical equations of motion of field theories
which are of relevance to  certain domains of physics is of great interest. One of these field theories
is the Abelian Higgs model whose importance embraces particle physics, condensed matter
physics and cosmology \cite{nielsen-olesen,witten,bogomolny,vilenkin-shellard,manton-sutcliffe}. 
The full fledged  Euler-Lagrange equations of motion of this model have so far
resisted attempts to solve them. This is because these are highly non-linear coupled second order partial differential
equations. In the literature, the classical studies  of the Abelian Higgs model are mostly dedicated 
to its topological properties. In this context, vortices have been identified and a great deal has been learned   
through numerical analyses (see \cite{klacka-et-al} for a sample). 
An account of the problem of constructing stationary topological and non-topological classical solutions  
in various field theories can be found in \cite{lee-pang, shnir,nugaev-et-al, volkov-radu}. 

\par
In order to reduce the degree of complexity of the equations of motion of the Abelian Higgs model
some simplifying strategies have to be adopted. In this area, the authors of ref.\cite{brihaye-et-al} explicitly constructed 
periodic sphaleron solutions (minima and saddle points of the energy functional) of the $(1+1)$-dimensional Abelian
Higgs model on a circle. They found some analytical solutions (for some special values of the Higgs mass) for the
the small perturbations (normal modes) around the sphaleron solutions. Their work is directly inspired 
by an earlier investigation by Manton and Samols \cite{manton-samols} who found sphaleron solutions in the scalar theory 
$\phi^4$  defined on a circle.
\par
In this note we have identified another situation where the equations of motion of the four-dimensional 
Abelian Higgs model become relatively simple. If the complex scalar field is parametrised as $\phi=\rho\, e^{i\theta}$ 
and we define the gauge invariant quantity $\widetilde A_\mu =A_\mu +\frac{1}{e} \dd_\mu\theta$, with $A_\mu$ being the 
gauge field, then imposing the constraint $\widetilde A_\mu \widetilde A^\mu=0$ renders the equations of motion 
tractable\footnote{Upon publication of the present work, I became aware that the authors of 
refs.\cite{canfora-1,canfora-2, canfora-3,canfora-4,canfora-5,canfora-6,canfora-7,canfora-8} have used, among others, the condition 
$A_\mu A^\mu =0$, to build analytic solutions in numerous field theories. I am greatful to Fabrizio Canfora for pointing out this to me.}. 
As a matter of fact, the equation of motion of the complex scalar field decouples and reduces to
the usual equation of motion  of  a $\phi^4$ scalar field theory which is known to possess kink solutions.  The problem is then brought 
to solving the gauge field equation in the presence of a kink background. 
\par
We have first solved the gauge field equation of motion when the scalar field $\rho$ takes the usual kink profile 
$\rho =  \pm\,v\,\tanh\left(p_\mu x^\mu + w_0\right)$, where $p_\mu$ is a space-like four vector. The gauge field
$A_\mu$ is, up to a gauge function, expressed in terms of associated Legendre functions and has two independent polarisations.
However, the equation of motion of a $\phi^4$ scalar field theory is also solved by the twelve Jacobi elleptic functions
(the kink solution is a very special case of this). Next, we solved the gauge field equation of motion when the 
scalar field  $\rho$ is represented by one of the Jacobi elleptic functions. Here we found that the 
gauge field is determined by means of  Heun functions. 

\par
In passing, we should mention that some analytical solutions involving gauge fields have been 
explicitly constructed in different contexts.  Brihaye \cite{brihaye}, in the case of the
$SU(2)$ Yang-Mills theory coupled to a triplet Higgs field, has found (for particular values of the ratio of the two coupling constants) solutions involving
the Jacobi elleptic functions.
The authors of ref.\cite{diakonos-et-al}, studying  a $(1+1)$-dimensional Abelian Higgs model,
have obtained  exact solutions (subject to the approximation that the two wells of  the Higgs potential are  deep). 
These were consequently used to build approximate analytical solution, in the form of oscillons and oscillating kinks,
for the dynamics of both the gauge and Higgs fields. Similar studies  can also be found in \cite{achellios-et-al,katsimiga}.
Along these lines, one finds in \cite{rozowsky-et-al} a numerical solutions corresponding to a kink (domain wall)
in a theory consisting of two interacting  complex scalars coupled to two independant gauge fields. The stability of this solution
was later analysed in \cite{george-volkas}.  
The partial relevance of all of these solutions to the present work is worth 
investigating.\footnote{I am greatful to an anonymous referee for bringing some of these
works to my attention.}  Finally, various analytical solutions involving a generalised Maxwell-Higgs models (theories with non-standard 
kinetic terms) have been obtained in \cite{bazeia,casana}. Nevertheless, we should insist on the fact that none of the
above mentioned solutions in \cite{brihaye-et-al,diakonos-et-al,achellios-et-al,katsimiga,rozowsky-et-al,george-volkas} is exact.
The closest study to our analyses is ref.\cite{brihaye-et-al} (in that it involves the Jacobi elleptic functions). Their
sphaleron solution (and the perturbation about it) to not fit in the caterory  $\widetilde A_\mu \widetilde A^\mu=0$  considered here.
 
\par
The paper is organised as follows: In the next section we briefly describe the Abelian Higgs model and 
lay out its simple classical solutions. In section three, we solve the equation of motion of the gauge field
in the background of a kink scalar field. We then review, in section four, the classical solution to the equation
of motion of a  $\phi^4$ scalar field theory in terms of the Jacobi elleptic functions (solution involving the  Weierstrass elliptic
function are presented in an appendix). These are used in section five
to build the solution to the gauge field equation of motion. Our main results are summarised in the last section.

\section{The Abelian Higgs model}

The Abelian Higgs model is described by the Lagrangian\footnote{The space-time coordinates
are $x^\mu=\left(x^0\,,\,x^1\,,\,x^2\,,\,x^3\right)=\left(x^0\,,\,\vec{x}\right)$ and the indices are raised 
and lowered with the metric $g_{\mu\nu}={\rm{diag}}\left(1\,,\,-1\,,\,-1\,,\,-1\right)$. 
A four-vector is $V^\mu=\left(V^0\,,\,V^1\,,\,V^2\,,\,V^3\right)=\left(V^0\,,\,\vec{V}\right)$.}
\bea
{\cal L}
= 
-\frac{1}{4}F_{\mu\nu}F^{\mu\nu}+
\left(\dd_\mu\phi^\star-ieA_\mu\phi^\star\right)
\left(\dd^\mu\phi+ieA^\mu\phi\right)
 -\frac{\lambda}{2}\left(\varphi^\star \varphi -v^2\right)^2
 \,\,\,.
\label{A-H-M}
\eea
Here $F_{\mu\nu} =  \dd_\mu A_\nu - \dd_\nu A_\mu$
is the field strength of the gauge field
$A_\mu $ and $\phi$ is the complex scalar field.
It is understood that the two parameters $v^2$ and $\lambda $ are both positive. 

\par
It is convenient for our purpose to  parametrise the complex scalar field $\phi$ as\footnote{One must be aware
that $\phi=\rho\, e^{i\theta}=\rho\, e^{i(\theta+2\pi N)}$ with $N\in \mathbb{Z}$. 
In this note we assume that $\dd_\mu \dd_\nu\theta - \dd_\nu \dd_\mu \theta=0$.}
\be
\phi=\rho\, e^{i\theta} \,\,\,\,.
\ee
\label{phi-para}
The Lagrangian (\ref{A-H-M}) becomes then
\bea
{\cal L} &=&
-\frac{1}{4}F_{\mu\nu}F^{\mu\nu}
+
e^2\rho^2\left(A_\mu +\frac{1}{ e} \dd_\mu\theta \right)
\left(A^\mu +\frac{1}{e} \dd^\mu\theta \right)
\nonumber \\
& +&\dd_\mu\rho\, \dd^\mu\rho\ -
\frac{\lambda}{2}\left(\rho^2 -v^2\right)^2
\,\,\,\,
\label{rho-theta}
\eea
and the local gauge symmetry is 
\bea
A_\mu \longrightarrow  A_\mu + \frac{1}{e}\dd_\mu\Lambda
\,\,\,\,\,\,\,,\,\,\,\,\,\,\,
\theta \longrightarrow \theta - \Lambda 
\,\,\,\,\,\,\,,\,\,\,\,\,\,\,
\rho \longrightarrow \rho 
\,\,\,\,\,\,.
\label{AGS-theta}
\eea
One could use this gauge freedom to set the field $\theta$ to zero.
Nevertheless, we will keep  $\theta$ through out.

\par
The equations of motion for the Abelian Higgs model are
\bea
\partial_\mu \widetilde  F^{\mu\nu} + 2 e^2\,\rho^2\,\widetilde A^\nu &=& 0 \,\,\,\,\,,
\label{eom1}
\\
\dd_\mu\dd^\mu \rho + {\lambda}\,\rho\left(\rho^2 -v^2\right)
-e^2\,\rho\, \widetilde A_\mu \widetilde A^\mu  
 &=&0\,\,\,\,\,,
\label{eom2}
\\
\dd_\mu \left(\rho^2 \widetilde A^\mu \right) &=& 0
\,\,\,\,\,.
\label{eom3}
\eea
We have defined the gauge invariant variable
\be
\widetilde A_\mu =A_\mu +\frac{1}{e} \dd_\mu\theta  \,\,\,\,.
\label{abelian-decomp}
\ee
and  $\widetilde F_{\mu\nu} = F_{\mu\nu} = \dd_\mu \widetilde A_\nu - \dd_\nu \widetilde A_\mu $.
The last equation (\ref{eom3}) corresponds to the field $\theta$  and  is also a consequence 
of (\ref{eom1}). Notice that we have expressed the equations of motion in terms
of the two gauge invariant variables $\widetilde A_\mu $ and $\rho$.

\par
The simplest known solution to the equations of motion is when the scalar field $\rho$ is
frozen at one of the two  minima of the potential energy. That is, $\rho^2=v^2$. In this case
the equations of motion become
 \bea
\partial_\mu \widetilde  F^{\mu\nu} + 2 e^2\,v^2\,\widetilde A^\nu &=& 0 \,\,\,\,\,,
\label{Ceom1}
\\
\widetilde A_\mu \widetilde A^\mu  
 &=& 0\,\,\,\,\,,
\label{Ceom2}
\\
\dd_\mu \widetilde A^\mu &=& 0
\,\,\,\,\,.
\label{Ceom3}
\eea
These equations have, for instance,  the plane wave solution
\bea
\widetilde A_\mu=\varepsilon_\mu\,cos\left(p_\nu x^\nu +w_0\right)\,\,\,\,\,\,\,\,\,\,,\,\,\,\,\,\,\,\,\,\,
\varepsilon_\mu \,\varepsilon^\mu = \varepsilon_\mu \,p^\mu =0
\,\,\,\,\,\,\,\,\,\,,\,\,\,\,\,\,\,\,\,\, p_\mu\,p^\mu =  2 e^2\,v^2 \,\,\,\,.
\eea
The polarisation vector $\varepsilon_\mu$  has two independent components for a given
wave vector $p_\mu$. The mass-shell relation  $p_\mu\,p^\mu =  2 e^2\,v^2$ is that of a 
massive particle with a positive mass squared. 

\par
The other known solution is the ''kink''  solution for which $\widetilde A_\mu = 0$. The equations
of motion come then to the single equation 
\be
\dd_\mu\dd^\mu \rho + {\lambda}\,\rho\left(\rho^2 -v^2\right)\,=\,0\,\,\,\,\,.
\ee
This is solved by
\bea
\rho =  \pm\,v\,\tanh\left(p_\mu x^\mu + w_0\right)\,\,\,\,\,\,\,\,\,\,,\,\,\,\,\,\,\,\,\,\,
p_\mu p^\mu =-\frac{\lambda v^2}{2} \,\,\,\,\,.
\label{kink-sol}
\eea
Here we have a mass-shell condition of a relativistic particle of negative mass squared.


In this note, we will look for a solution which satisfies the gauge invariant condition 
\be
\widetilde A_\mu \widetilde A^\mu=0\,\,\,\,\,.
\label{AA}
\ee
The equations of motion reduce then to 
\bea
\partial_\mu \widetilde  F^{\mu\nu} + 2 e^2\,\rho^2\,\widetilde A^\nu &=& 0 \,\,\,\,\,,
\label{eom111}
\\
\dd_\mu\dd^\mu \rho + {\lambda}\,\rho\left(\rho^2 -v^2\right)
 &=&0\,\,\,\,\,,
\label{eom222}
\\
\dd_\mu \left(\rho^2 \widetilde A^\mu \right) &=& 0
\,\,\,\,\,.
\label{eom333}
\eea
We notice that the equation of motion for the scalar field $\rho$ decouples from the rest.
We will report here on a non-trivial solution obeying the condition (\ref{AA}).

\section{A gauge field corresponding to the ``kink'' scalar field}

\par
The second equation (\ref{eom222}) admits the ''kink'' solution as written in (\ref{kink-sol}).
We would like here to find the gauge field corresponding to it.
We assume the following form for the gauge field $\widetilde A_\mu$ :
\be
\widetilde A_\mu\left(w\right)=\varepsilon_\mu\,h\left(w\right) \,\,\,\,\,.
\label{h}
\ee
Here (and in the rest of the paper)  we will use the notation 
\be
w=p_\mu x^\mu + w_0\,\,\,\,\,\,\,\,\,\,,\,\,\,\,\,\,\,\,\,\,
p^2 = p_\mu p^\mu
\ee
with $w_0$  a constant.  The four-vector $p_\mu$ obeyes the mass-shell relation
$p^2 =-\frac{\lambda v^2}{2}$.
The polarisation vector $\varepsilon_\mu$ is required to satisfy
\be
\varepsilon_\mu\,\varepsilon^\mu\, = p_\mu \,\varepsilon^\mu\,=0
\,\,\,\,\,.
\ee
Equations (\ref{AA}) and  (\ref{eom333}) are then automatically obeyed.

\par
Substituting  (\ref{h}) into  (\ref{eom111}) results in the differential equation
\be
\frac{d^2 h}{dw^2}\left(w\right) -\frac{4e^2}{\lambda}\,\tanh^2\left(w\right)\,h\left(w\right)=0
\,\,\,\,\,.
\label{ddh}
\ee
By the change of variables
\be
z=\tanh\left(w\right)
\ee
one transforms the differential equation  (\ref{ddh}) into
\bea
\left(1-z^2\right)\,\frac{d^2 h}{dz^2}\left(z\right)
-2z\,\frac{d h}{dz}\left(z\right)+\left[l\left(l+1\right)-\frac{m^2}{\left(1-z^2\right) }\right]h\left(z\right)=0
\,\,\,\,\,.
\label{Legendre}
\eea
This differential equation is known as associated Legendre equations \cite{gradshteyn,abramowitz,nist}. 
It is, in general, singular at the points $z=\pm 1\,,\,\pm\infty$.
In the case at hand, the variable $z$ is real and lies  in the domain $\left[-1\,,\,+1\right]$ and 
the real constants $l$ and $m$ are defined as\footnote{Greek indices $\nu$ and $\mu$ are usually used instead of $l$ and $m$.} 
\bea
l=-\frac{1}{2} \pm \sqrt{\frac{1}{4}+\frac{4e^2}{\lambda}}
\,\,\,\,\,\,\,\,\,\,\,\,\,\,\,,\,\,\,\,\,\,\,\,\,\,\,\,\,\,\,
m^2= \frac{4e^2}{\lambda}
\,\,\,\,\,\,.
\label{l-m}
\eea
The general solution  to  (\ref{Legendre}) is
\bea
h\left(z\right)=a\,P^m_l\left(z\right) + b\,Q^m_l\left(z\right)\,\,\,\,\,\,,
\eea
where $P^m_l\left(z\right)$, $Q^m_l\left(z\right)$ are associated Legendre functions \cite{gradshteyn,abramowitz,nist}
of the first and second kind, respectively.
The two constants $a$ and $b$ are arbitrary.

\par
When $z$ is real and belonging to the interval $\left[-1\,,\,+1\right]$, the associated  Legendre function 
$P^m_l\left(z\right)$ is real and is expressed  as \cite{gradshteyn,abramowitz,nist}
\bea
P^m_l\left(z\right)=\frac{1}{\Gamma\left(1-m\right)}\left(\frac{1+z}{1-z}\right)^{\frac{m}{2}}
F\left(-l,l+1;1-m;\frac{1-z}{2}\right)\,\,\,\,\,\,,
\eea
where $F\left(a,b;c;x\right)$ is the hypergeometric function and $\Gamma\left(\nu\right)$ is the gamma 
function. Similarly, the associated  Legendre function 
$Q^m_l\left(z\right)$ is given by
\bea
Q^m_l\left(z\right)=\frac{\pi}{2\sin\left(m\pi\right)}\left[\cos\left(m\pi\right)\,P^m_l\left(z\right)
- \frac{\Gamma\left(l+m+1\right)}{\Gamma\left(l-m+1\right)}\,P^{-m}_l\left(z\right)\right]\,\,\,\,\,\,.
\eea
\par
To summarise, a particular solution to the equations of motion of the Abelian
Higgs model (\ref{A-H-M})  is given by
\bea
\phi  &=& \pm\,v\,\tanh\left(w\right)\,e^{i\theta}\,\,\,\,\,\,,
\nonumber \\
A_\mu &=& - \frac{1}{e}\,\dd_\mu\theta + \varepsilon_\mu
\left[a\,P^m_l\left(\tanh\left(w\right)\right) + b\,Q^m_l\left(\tanh\left(w\right)\right)\right]\,\,\,\,\,\,,
\nonumber \\
p^2 &=& -\frac{\lambda v^2}{2}\,\,\,\,\,\,,\,\,\,\,\,\,
\varepsilon_\mu\,\varepsilon^\mu\, = p_\mu \,\varepsilon^\mu\,=0
\,\,\,\,\,\,.
\label{sol1}
\eea
The parameters $l$ and $m$ are given (\ref{l-m}). 
The field $\theta\left(x\right)$ is any arbitrary smooth function (obeying $\dd_\mu\dd_\nu \theta-\dd_\nu\dd_\mu \theta=0$). 
The gauge field $A_\mu$
has two independent polarisations for a given vector $p_\mu$. Furthermore, 
the solution carries both electric and magnetic fields.


\section{Solutions to \texorpdfstring{$\phi^4$}{phi} in terms of the Jacobi elliptic functions}

The equation of motion for a phi-to-the-four scalar field theory is given by
\bea
\dd_\mu\dd^\mu \rho + {\lambda}\,\rho\left(\rho^2 -v^2\right)
 =0\,\,\,\,\,.
\label{phi-four-1}
\eea
We will look for solution  which depend on the single variable
$
w=p_\mu x^\mu +w_0$. That is, $\rho=\rho\left(w\right)$.
The equation of motion becomes then
\bea
\frac{\text{d}^2\rho}{\text{d}w^2} + \frac{\lambda}{p^2}\,\rho\left(\rho^2 -v^2\right)
 &=& 0\,\,\,\,\,\,.
\label{phi-four-2}
\eea
Notice that any solution to this equation is  obviously subject to the following remark:
\bea
\text{if}\,\,\, \rho\left(w\,,\,p^2\right)\,\,\,\text{is a solution then } \,\,\,
 \rho\left(s\,w\,,\,\frac{p^2}{s^2} \right)\,\,\,\text{is also a  solution }\,\,\,,
\label{remark}
\eea
where $s$ is a constant. Therefore, the mass-shell relation, $p^2$,  will be determined up
to a constant.

\par
It is well-known that equation (\ref{phi-four-2}) is  solved by the twelve Jacobi elliptic functions \cite{gradshteyn,abramowitz,nist}. Indeed, 
The scalar field $\rho(w)$ satisfies  the  first order differential equation 
\bea
\left(\frac{\text{d}\rho}{\text{d}w} \right)^2= -\frac{\lambda }{2p^2}\,\rho^2\left(\rho^2-2v^2\right) 
+av^2
\,\,\,\,\,
\eea
with $av^2$ a constant of integration.
By writing 
\bea
\rho\left(w\right)=v\,\varepsilon\, y\left(w\right)\,\,\,\,\,,
\eea
where $\varepsilon$ is a constant, one obtains 
the first order differential equation 
\bea
\left(\frac{\text{d}y}{\text{d}w} \right)^2= -\frac{\lambda v^2 \varepsilon^2}{2p^2}\,y^4 + 
\frac{\lambda v^2 }{p^2}\,y^2 + \frac{a}{\varepsilon^2}\,\,\,\,\,.
\label{dy2}
\eea
The twelve Jacobi elliptic functions are 
obtained as solutions to this last equation \cite{gradshteyn,abramowitz,nist} for different choices of the three constants
$p^2$,  $\varepsilon^2$ and $a$ (and boundary conditions). The table in Appendix B gather these values.
{}For instance, the
solution given in terms of the ``sine'' Jacobi elliptic function ${\text{sn}}{(x\,,\,m)}$ corresponds to  the choice
\bea
a &=& \varepsilon^2 =  \frac{2m^2}{1+m^2} \,\,\,\,,
\nonumber \\
p^2 &=& -\frac{\lambda v^2}{1+m^2}  \,\,\,\,
\eea
and the differential equation (\ref{dy2}) takes the form
\bea
\left(\frac{\text{d}y}{\text{d}w} \right)^2= 
\left(1-y^2\right)\left(1-m^2\,y^2\right)
\,\,\,\,\,.
\label{sn-param}
\eea
The general solution \cite{gradshteyn,abramowitz,nist} to this last equation is $y\left(w\right)={\text{sn}}{(w+d\,,\,m)}$, where
$d$ is an arbitrary constant. 
Hence our scalar field $\rho$ is given by 
\bea
\rho\left(w\right)=\pm v\,\sqrt{\frac{2m^2}{1+m^2}}\,\,{\text{sn}}{(w+d\,,\,m)}
\,\,\,\,\,\,,\,\,\,\,\,\, 
p^2=-\frac{\lambda v^2}{1+m^2}  \,\,\,\,.
\label{sn-sol}
\eea
The parameter $m$ must be different from zero here.
The mass-shell relation $p^2=-\frac{\lambda v^2}{1+m^2}$  is that of a particle with 
negative mass squared. 
It is worth mentioning that the ``kink'' solution (\ref{kink-sol}) corresponds to $m=1$
as 
\be
{\text{sn}}{(w+d\,,\,1)}=\tanh\left(w+d\right) \,\,\,\,.
\ee

\section{The general solution to the gauge field equation}

The gauge field, $A_\mu=\varepsilon_\mu\,h\left(w\right)$ with $\varepsilon_\mu\varepsilon^\mu = p_\mu\varepsilon^\mu=0$,
 is determined by the differential equation
\be
\frac{\text{d}^2 h}{\text{d}w^2}\left(w\right) +\frac{2e^2}{p^2}\,\rho^2\left(w\right)\,h\left(w\right)=0
\,\,\,\,\,.
\label{ddh-1}
\ee
{}For the moment, the expression of $p^2$ is not fixed. This allows one to treat all possible
mass-shell conditions at the same time.
\par
Let us assume  that
\be
h\left(w\right)=h\left(Z\right)\,\,\,\,\,,\,\,\,\,\,Z=\frac{1}{\tau^2}\,\rho^2\left(w\right)
\,\,\,\,\,,
\ee
where $\tau$ is a constant to be properly chosen later.
Then by using the fact that the field $\rho\left(w\right)$ satisfies the two equations
\bea
\frac{\text{d}^2\rho}{\text{d}w^2} &=&- \frac{\lambda}{p^2}\,\rho\left(\rho^2 -v^2\right)
\,\,\,\,\,,
\nonumber \\
\left(\frac{\text{d}\rho}{\text{d}w} \right)^2 &=& -\frac{\lambda }{2p^2}\,\rho^2\left(\rho^2-2v^2\right) 
+av^2
\,\,\,\,\, 
\eea
we arrive at the differential equation
\bea
\frac{\text{d}^2 h}{\text{d}Z^2}
+\left(\frac{\gamma}{Z}-\frac{\delta}{1-Z}-\frac{\epsilon k^2}{1-k^2Z}\right)\frac{\text{d} h}{\text{d}Z}
+\frac{\left(s+\alpha\beta k^2 Z\right)}{Z\left(1-Z\right)\left(1-k^2 Z\right)}\,h=0
\,\,\,\,\,.
\label{Heuneq}
\eea
The different constants are given by
\bea
s &=& 0\,\,\,\,\,,
\nonumber \\
\delta &=& \gamma =\epsilon= \frac{1}{2}\,\,\,\,\,,
\nonumber \\
\alpha  &=& \alpha_\pm = \frac{1}{4}
\left(1 \pm \sqrt{1+16\,\frac{e^2}{\lambda}}\right)
\,\,\,\,\,,
\nonumber \\
\beta  &=& \beta_\mp = \frac{1}{4}
\left(1 \mp \sqrt{1+16\,\frac{e^2}{\lambda}}\right)
\,\,\,\,\,,
\nonumber \\
\tau^2  &=& \frac{2  k^2}{\left(1+k^2\right)}\,v^2
\,\,\,\,\,.
\label{Hn-param}
\eea
They satisfy the relation
\be
\gamma +\delta +\epsilon = \alpha_\pm + \beta_\mp +1 \,\,\,\,\,.
\ee
We also have the mass-shell relation 
\bea
p^2 &=& -\frac{2\lambda v^2 k^2}{a\left(1+k^2\right)^2}
\,\,\,\,\,.
\label{p2}
\eea
The constant $a$ will be fixed later.
\par
The equation (\ref{Heuneq}) is a  Fuchsian differential equation and its  
general solution is given by 
\bea
h\left(Z\right)  &=& C_1\, \text{Hn}\left(k^2,0;\alpha_+,\beta_-,\frac{1}{2},\frac{1}{2};Z\right)
 + C_2 \,\text{Hn}\left(k^2,0;\alpha_-,\beta_+,\frac{1}{2},\frac{1}{2};Z\right)
\,\,\,\,\,,
\nonumber \\
Z &=& \frac{1}{\tau^2}\,\rho^2\left(w\right) 
= \frac{\left(1+k^2\right)}{2  k^2}\,\frac{1}{v^2}\, \rho^2\,\left(w\right) 
\,\,\,\,\,,
\label{h(Z)}
\eea
where $\text{Hn}\left(k^2,s;\alpha,\beta,\gamma,\delta;z\right)$ is the Heun function\footnote{Sometimes the notation
$a=1/k^2$ and $q=-s/k^2$ is used. A useful note on Heun's functions and further references can be found in \cite{valent}.} 
and $C1$ and $C2$ are two arbitrary constants. It is important to emphasise at this point that 
the scalar field $\rho\left(w\right)$ is any solution to the equation of motion (\ref{phi-four-2}),
that is,  any of the twelve Jacobi elliptic functions.

\par
When, for instance,  the  scalar field  $\rho\left(w\right)$ is expressed in terms of 
the Jacobi elliptic function
$ \text{sn}\left(w+d,m\right)$ in  (\ref{sn-sol})  then the expression of $p^2$ there  has to match that 
written in  (\ref{p2}).  
This leads to the identifications
\bea
a = \frac{2 k^2\left(1+m^2\right)}{\left(1+k^2\right)^2}=
\frac{2 m^2}{\left(1+m^2\right)}
\,\,\,\,\,\,\,\, \Longrightarrow \,\,\,\, k^2=m^2 \,\,\,\,\,\,\,\,
\text{or}  \,\,\,\,\,\,\,\, k^2=\frac{1}{m^2}
\,\,\,\,\,.
\label{k=m}
\eea
If we choose $k=m$, for example,  then,  using (\ref{sn-sol}) and (\ref{h(Z)}), the  two fields of the Abelian
Higgs model (\ref{A-H-M})  are  found to be  given by
\bea
\phi  &=& \pm v\,\sqrt{\frac{2m^2}{1+m^2}}\,{\text{sn}}{(w+d\,,\,m)} \,e^{i\theta}\,\,\,\,\,\,,
\nonumber \\
A_\mu &=& - \frac{1}{e}\,\dd_\mu\theta + \varepsilon_\mu \left[
 C_1\, \text{Hn}\left(m^2,0;\alpha_+,\beta_-,\frac{1}{2},\frac{1}{2}; {\text{sn}}^2{(w+d\,,\,m)}  \right)
\right.
\nonumber \\
 &+&  C_2 \,\text{Hn}\left(m^2,0;\alpha_-,\beta_+,\frac{1}{2},\frac{1}{2}; {\text{sn}}^2{(w+d\,,\,m)}  \right)
 \bigg]\,\,\,\,\,,
\nonumber \\
p^2 &=& -\frac{\lambda v^2}{1+m^2}  \,\,\,\,\,\,\,\,, \,\,\,\,\,\,\,\, 
\varepsilon_\mu\,\varepsilon^\mu\, = p_\mu \,\varepsilon^\mu\,=0
\,\,\,\,\,\,. 
\label{sol-Jacobi}
\eea
The two constants $\alpha_\pm$ and $\beta_\pm$ depend on  the Abelian Higgs parameters 
and are listed in (\ref{Hn-param}).
The field  $\theta\left(x\right)$ is  arbitrary. 

\par
Finally, when $m=1$ one has ${\text{sn}}{(w+d\,,\,1)}=\tanh\left(w+d\right)$, and the particular solution 
given in (\ref{sol1}) is recovered by converting the Heun functions into 
hypergeometric functions with the help of the relation \cite{valent}
\bea
\left\{
\begin{array}{l}
\text{Hn}\left(1,s;\alpha,\beta,\gamma,\delta;z\right)
=\left(1-z\right)^r F\left(r+\alpha,r+\beta;\gamma;z\right)\,\,\,\,,\\
\\
r=\xi -\sqrt{\xi^2 -\alpha\beta-s}\,\,\,\,\,\,,\,\,\,\,\,\,\,
\xi=\frac{1}{2}\left(\gamma-\alpha-\beta\right)\,\,\,\,\,.
\end{array} \right.
\eea

\par
\begin{figure}[thp]
\begin{center}
\includegraphics[scale=0.2]{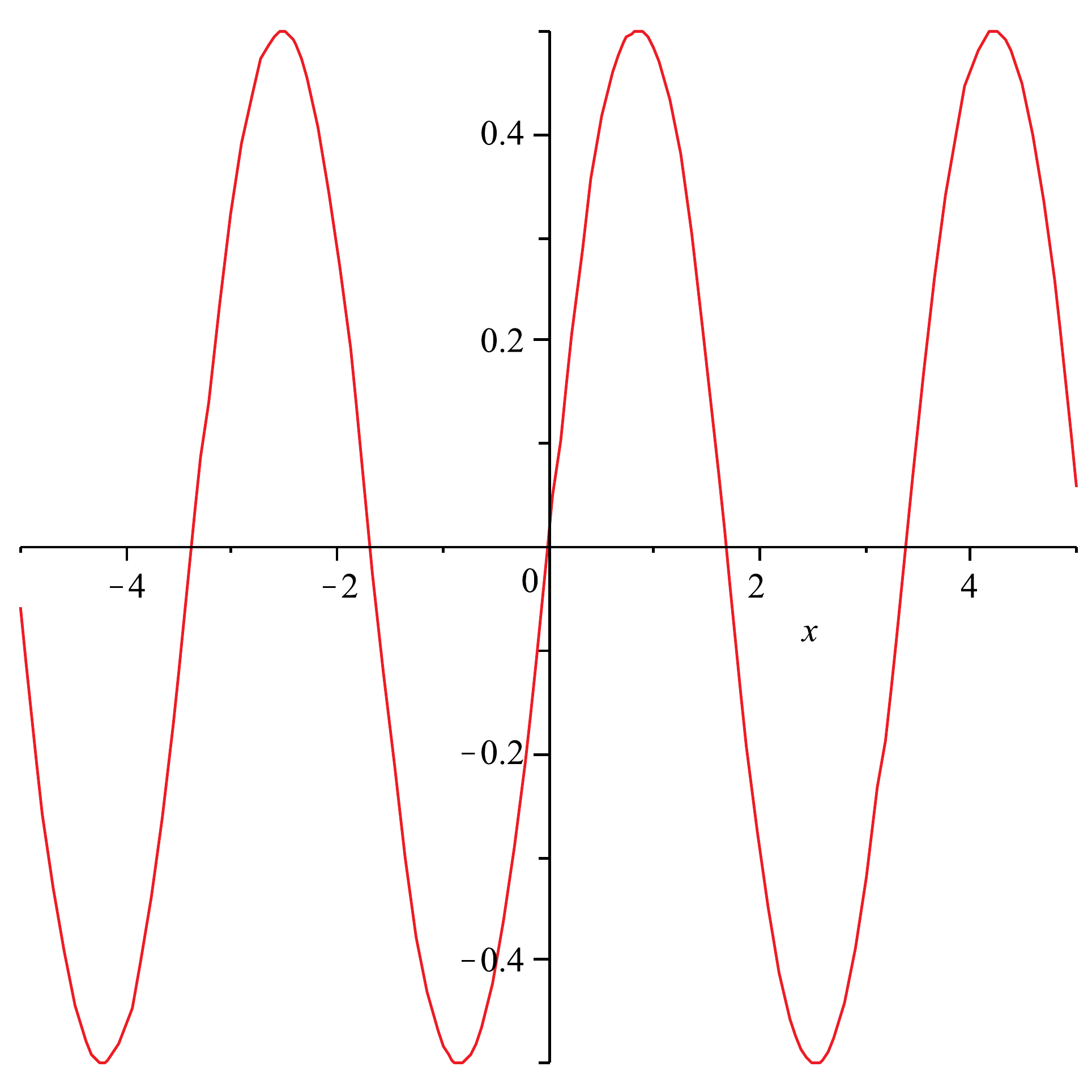}
\end{center}
\caption[]{A sketch of the Jacobi elliptic function ${\text{sn}}{(x\,,\,2)}$.}
\label{B}
\end{figure}
\begin{figure}[thp]
\begin{center}
\includegraphics[scale=0.2]{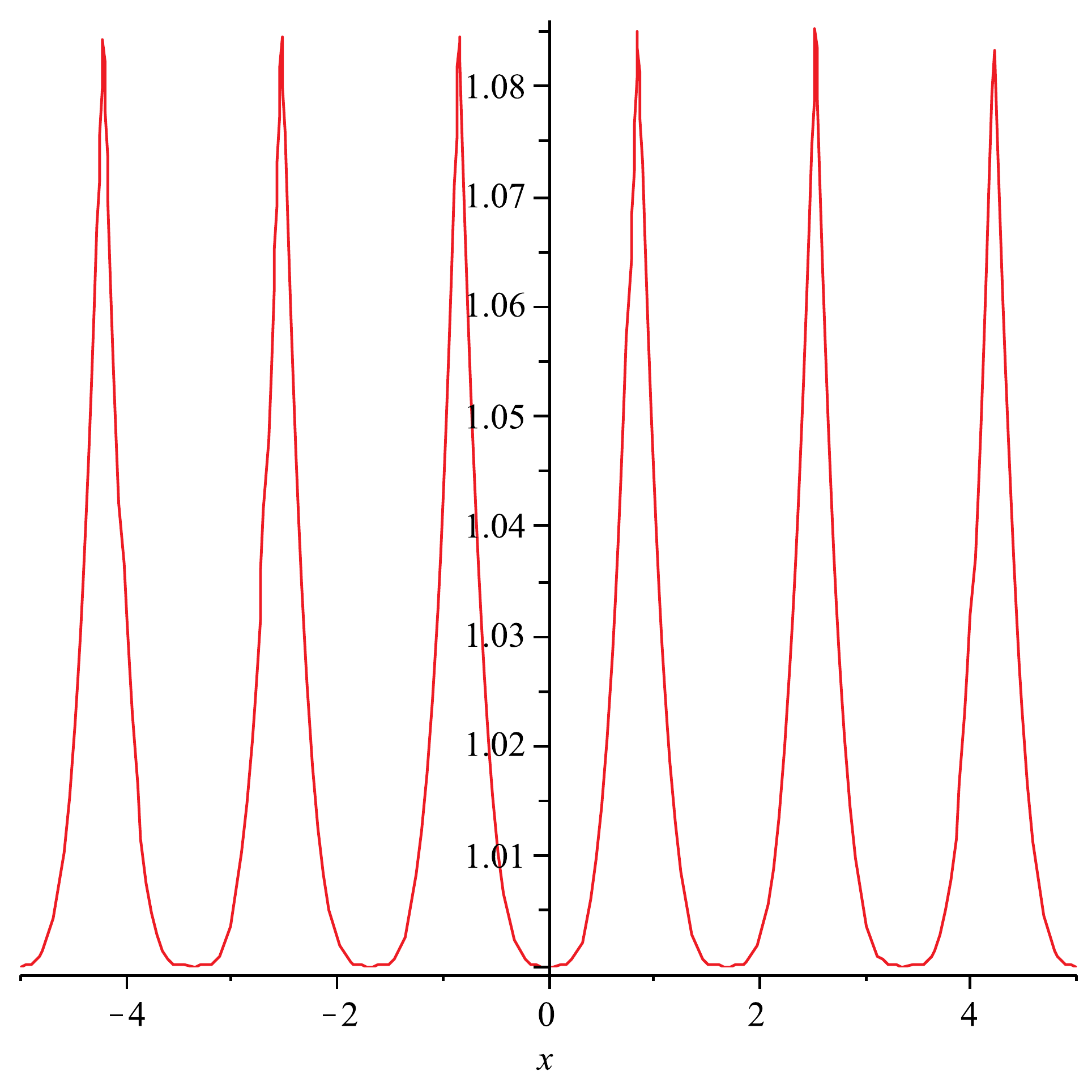}
\end{center}
\caption[]{A sketch of the Heun function $\text{Hn}\left(4,0;\frac{3}{4},-\frac{1}{4},\frac{1}{2},\frac{1}{2}; {\text{sn}}^2{(x,2)}\right)$ for
$\sqrt{1+16\frac{e^2}{\lambda}}=2$.}
\label{BB}
\end{figure}

We have represented in Figure \ref{B} the Jacobi elliptic function ${\text{sn}}{(x,2)}$
and in Figure \ref{BB} the Heun function $\text{Hn}\left(4,0;\frac{3}{4},-\frac{1}{4},\frac{1}{2},\frac{1}{2}; {\text{sn}}^2{(x,2)}\right)$.

\section{Conclusion}

We have presented classical solutions to the equations of motion   of the Abelian Higgs model (\ref{A-H-M}).
The complex scalar field $\phi$ and the gauge field $A_\mu$ are given by
\bea
\phi  &=& \pm v\,\varepsilon\,{\text{pq}}{\left(w+d\,,\,m\right)} \,e^{i\theta}\,\,\,\,\,\,,
\nonumber \\
A_\mu &=& - \frac{1}{e}\,\dd_\mu\theta + \varepsilon_\mu \bigg[
 C_1\, \text{Hn}\left(k^2,0;\alpha_+,\beta_-,\frac{1}{2},\frac{1}{2};Z \right)
\nonumber \\
 &+&  C_2 \,\text{Hn}\left(k^2,0;\alpha_-,\beta_+,\frac{1}{2},\frac{1}{2};Z \right)
 \bigg]\,\,\,\,\,,
\nonumber  \\
Z &=& 
\varepsilon^2\, \frac{\left(1+k^2\right)}{2  k^2}\,{\text{pq}}^2{\left(w+d\,,\,m\right)} 
\,\,\,\,\,,
\nonumber \\
w &=& p_\mu x^\mu +w_0 \,\,\,\,\,\,\,\,, \,\,\,\,\,\,\,\, 
\varepsilon_\mu\,\varepsilon^\mu\, = p_\mu \,\varepsilon^\mu\,=0
\,\,\,\,\,\,. 
\label{sol-con}
\eea
Here ${\text{pq}}{\left(w+d\,,\,m\right)}$, with   ${\text{pq}}$ any pair of the letters (c,d,n,s), is one of the 
twelve Jacobi elliptic functions and $\text{Hn}\left(k^2,s;\alpha,\beta,\gamma,\delta;z\right)$ is the Heun function. 
{} We have listed in table \ref{tab1-1} the values $p^2$ (the mass-shell relation) and 
$\varepsilon$ for each function ${\text{pq}}{\left(w+d\,,\,m\right)}$. Table \ref{tab1-1} gives also the expression
of $k^2$ in terms of the parameter $m$ entering the  Jacobi elliptic function ${\text{pq}}{\left(w+d\,,\,m\right)}$.
Finally, $\alpha_\pm = \frac{1}{4}
\left(1 \pm \sqrt{1+16\,\frac{e^2}{\lambda}}\right)$
and
$\beta_\mp = \frac{1}{4}
\left(1 \mp \sqrt{1+16\,\frac{e^2}{\lambda}}\right)$.

\par
Notice that the differential equation (\ref{ddh-1}) is linear in $h(w)$. Therefore, the gauge field $A_\mu$ is in fact
a linear combination of all the polarisation vectors $\varepsilon_\mu$ satisfying $\varepsilon_\mu\,\varepsilon^\mu\, = p_\mu \,\varepsilon^\mu\,=0$.
Finally, the solution (\ref{sol-con}) could have been expressed in terms of the Weierstrass elliptic function as shown in Appendix A.

\par
There are two immediate questions that one might ask regarding the solutions found in this article. The first 
regards their physical relevance. Unfortunately, our solution cannot find applications in condensed  matter physics. This is because 
the 'trick' used to decouple the complex scalar field equation of motion
does not apply in the non-relativistic version of the Abelian Higgs model (Ginzburg-Landau theory, gauged non-linear Schr\"odinger, London theory, $\dots$).
The equivalent of    $\widetilde A_\mu \widetilde A^\mu=0$  would be 
$A_0=c \vec A^2$,  $c$ being a constant. However, this is not compatible with Maxwell's equations. It is though plausible that our solution
might be of use in a theory of gravity coupled to the Abelian Higgs model. This is currently under investigation.

\par
The second (hard) question concerns the stability of the solution presented here. Our solution is not protected by 
any topological argument (unlike the vortex solution) and therefore is expected to be unstable. 
The only way to set one's mind at rest is by carrying a perturbation expansion around our exact solutions.
This is done through  the substitution\footnote{Our solution can be make
static by setting $p_0=0$ in the variable $w=p_\mu x^\mu +w_0$.} (in the spirit of refs.\cite{brihaye-et-al,george-volkas})
\bea
\rho &\longrightarrow & \rho\left(w\right)  + \kappa\left(w\right)\,e^{i\omega_1t}
\nonumber \\
\widetilde A_\mu  &\longrightarrow & \widetilde A_\mu\left(w\right) + a_\mu \left(w\right)\,e^{i\omega_2t}
\,\,\,.
\label{sub}
\eea
Here $\rho\left(w\right)$ and $\widetilde A_\mu\left(w\right)$ are solutions to the equations of motion of the Abelian Higgs model.
The perturbations $\kappa\left(w\right)$ and  $a_\mu \left(w\right)$ are accompanied by their normal modes represented by
the two constants $\omega_1$ and  $\omega_2$, respectively. If  $\omega_1$ and  $\omega_2$ are real then we have an oscillatory regime
and the solution  is stable. The next step is to plug the above substitution into the full equations of motion (\ref{eom1}),  (\ref{eom2})
and  (\ref{eom3}). To first order in the perturbation, one obtains then  the differential equations which, in principle,  allow one to determine
the normal modes. The differential equations obtained here involve the Jacobi elleptic functions and Heun functions and are, at the moment, 
difficult to analyse.  

\par
It might be easier, though,  to choose a perturbation which respects the constraint $\widetilde A_\mu \widetilde A^\mu=0$.
This can be achieved for instance by writing $a_\mu \left(w\right)= v_\mu g\left(w\right)$ and
demanding that $v_\mu v^\mu= \varepsilon_\mu v^\mu =0$, where $\varepsilon_\mu$ is the polarisation vector of $\widetilde A_\mu\left(w\right)$.
The substitution (\ref{sub}) is then carried out in  the simpler equations of motion (\ref{eom111}),  (\ref{eom222})
and  (\ref{eom333}).  We hope to report on the progress in this case in the near future.

\par
As a last remark, we mention that our solution is easily adapted to the case of a $(1+1)$-dimensional Abelian Higgs model
and the complex scalar field $\phi$ could be compactified on a circle of length $L$. This would make close contact with the
works in \cite{brihaye-et-al,manton-samols,liang-et-al} and might be of help in the determination of the  normal modes.

\appendix

\newpage

\begin{center}
{\bf APPENDICES}
\end{center}

\section{Solutions to \texorpdfstring{$\phi^4$}{phi}  in terms of the  Weierstrass elliptic function}

The Weierstrass elliptic function is a general solution to the first order
differential equation \cite{gradshteyn,abramowitz,nist} (see also \cite{pastras} for some lectures on
the suject)
\bea
\left(\frac{\text{d}\wp}{\text{d}z} \right)^2 &=& 4\,\wp^3\left(z\right) -g_2\,\wp\left(z\right) -g_3
\nonumber \\
 &=& 4\left(\wp-e_1\right)\left(\wp-e_2\right)\left(\wp-e_3\right)
\,\,\,\,.
\eea
The constants $g_2$ and $g_3$ are known as the lattice invariants
and $e_i\,\,,i=1,2,3$ are the roots of the Weierstrass normal cubic
equation $4z^3 -g_2z-g_3=0$.

\par
In order to obtain a Weierstrass differential equation in the case of $\phi^4$ theory,  we start from  equation  
(\ref{dy2})
\bea
\left(\frac{\text{d}y}{\text{d}w} \right)^2= -\frac{\lambda v^2 \varepsilon^2}{2p^2}\,y^4 + 
\frac{\lambda v^2 }{p^2}\,y^2 + \frac{a}{\varepsilon^2}
\eea
and multiply both sides with $4y^2$ to get
\bea
\left(\frac{\text{d}f}{\text{d}w} \right)^2 &=&  -\frac{2\lambda v^2 \varepsilon^2}{p^2}\,f^3 + 
\frac{4\lambda v^2 }{p^2}\,f^2 + \frac{4a}{\varepsilon^2}\,f\,\,\,\,,
\nonumber\\
f\left(w\right) &\equiv& y^2\left(w\right)\,\,\,\,.
\eea
The next step is to get rid of the term proportional to  $f^2$. This is achieved by 
the change of functions
\be
f\left(w\right)=g\left(w\right)+\frac{2}{3}\frac{1}{\varepsilon^2}\,\,\,.
\ee
The resulting differential equation is given by
\bea
\left(\frac{\text{d}g}{\text{d}w} \right)^2 &=&  -\frac{2\lambda v^2 \varepsilon^2}{p^2}\,g^3  
+\frac{4}{\varepsilon^2}\left(\frac{2}{3}\frac{\lambda v^2}{ p^2}  
+ {a}\right)g
 + \frac{8}{3} \frac{1}{\varepsilon^4}\left(\frac{4\lambda v^2}{9 p^2} + {a}\right) \,\,\,\,.
\eea
In order to obtain a Weierstrass differential equation, we choose the constant $\varepsilon$ such that
\be
p^2=-\varepsilon^2\,\frac{\lambda v^2}{2}\,\,\,.
\ee
This leads to
\bea
\left(\frac{\text{d}g}{\text{d}w} \right)^2 &=&  4\,g^3  
+ \frac{4}{\varepsilon^4}\left({\varepsilon^2}\,a -\frac{4}{3}\right)\,g
 + \frac{8}{3} \frac{1}{\varepsilon^6}\left({\varepsilon^2}\,a-\frac{8}{9}\right) \,\,\,\,,
\nonumber \\
&=& 4
\left[g- \frac{1}{\varepsilon^2}\left(\frac{1}{3} +\sqrt{1-{\varepsilon^2}\,a}\right)\right]
\left[g- \frac{1}{\varepsilon^2}\left(\frac{1}{3} -\sqrt{1-{\varepsilon^2}\,a}\right)\right]
\left[g+ \frac{2}{3} \frac{1}{\varepsilon^2}\right]
 \,\,\,\,.
\nonumber \\
\eea
The solution to this differential equation is given by the  Weierstrass
elliptic function
\bea
g\left(w\right)=\wp\left(w+d\,;\,-\frac{4}{\varepsilon^4}\left({\varepsilon^2}\,a -\frac{4}{3}\right)\,,\,
-\frac{8}{3} \frac{1}{\varepsilon^6}\left({\varepsilon^2}\,a-\frac{8}{9}\right)\right)\,\,\,\,,
\eea
where $d$ is a constant. Recalling that   $\rho\left(w\right)=v\varepsilon y\left(w\right)$, 
the square of our scalar field $\rho\left(w\right)$ is finally  given by
\bea
\rho^2\left(w\right) &= & v^2\left[\frac{2}{3}+\varepsilon^2\,
\wp\left(w+d\,;\,-\frac{4}{\varepsilon^4}\left({\varepsilon^2}\,a -\frac{4}{3}\right)\,,\,
-\frac{8}{3} \frac{1}{\varepsilon^6}\left({\varepsilon^2}\,a-\frac{8}{9}\right)\right)\right]
\,\,\,\,\,\,\,\,\,
\nonumber \\
&= & v^2\left[\frac{2}{3}+
\wp\left(\frac{1}{\varepsilon}(w+d)\,;\,- {4} \left({\varepsilon^2}\,a -\frac{4}{3}\right)\,,\,
-\frac{8}{3} \left({\varepsilon^2}\,a-\frac{8}{9}\right)\right)\right]
\,\,\,\,\,,\,\,\,\,
\nonumber \\
p^2 &=& -\varepsilon^2\,\frac{\lambda v^2}{2}
\,\,\,.
\label{Weier-sol}
\eea
In the last equality we have used the homogeneity relation \cite{gradshteyn,abramowitz,nist,pastras}
\be
\wp\left(z;g_2,g_3\right)= \mu^{-2}\wp\left( \mu^{-1} z; \mu^{4}g_2, \mu^{6}g_3\right)\,\,\,\,.
\ee

\par
Therefore we could have used the Weierstrass elliptic function (instead of the Jacobi elliptic functions)  
to express the solution to the Abelian Higgs model given in (\ref{sol-con}).
In this case, 
the identification of the two expressions of  $p^2$ in  (\ref{Weier-sol}) and 
(\ref{p2}) leads to
\be
{\varepsilon^2}\,a =\frac{4 k^2 }{\left(1+k^2\right)^2}
\,\,\,\, .
\ee
Finally, we should mention that  the Weierstrass elliptic function could be converted into Jacobi elliptic functions
\cite{gradshteyn,abramowitz,nist,pastras}.

\section{The twelve Jacobi elliptic functions Solutions to \texorpdfstring{$\phi^4$}{phi} }

The table below summarises the solutions to the equation 
\bea
\left(\frac{\text{d}y}{\text{d}w} \right)^2= -\frac{\lambda v^2 \varepsilon^2}{2p^2}\,y^4 + 
\frac{\lambda v^2 }{p^2}\,y^2 + \frac{a}{\varepsilon^2}\,\,\,\,\,
\label{dy2-1}
\eea
and gives the corresponding values of the parameters $p^2$ (the mass-shell relation), $\varepsilon^2$ and $a$
(see also  \cite{khaled,turks}
and  \cite{frasca} for some particular cases).
\begin{center}
\begin{table}[ht]
\begin{tabular}{| l| l| l| l| l|l| } 
 \hline
$p^2$ & $\varepsilon^2$ & $\frac{a}{\varepsilon^2}$ & $ y\left(w\right)$ & $\rho\left(w\right)$ & $k^2\,\,\text{or}\,\,\frac{1}{k^2} $  \\  
\hline

$-\frac{\lambda v^2}{1+m^2}$ &  $\frac{2m^2}{1+m^2}$   & $1$ & $ \text{sn}\left(w,m\right)$ 
& $\pm\,v\,\varepsilon\, \text{sn}\left(w+d,m\right)$ 
 & $m^2$ \\ 
\hline

 $-\frac{\lambda v^2}{1-2m^2}$ & $- \frac{2m^2}{1-2m^2}$   & $1-m^2$ & $ \text{cn}\left(w+d,m\right)$ &  
$\pm\,v\,\varepsilon\, \text{cn}\left(w+d,m\right)$ 
& $\frac{m^2-1}{m^2}$  \\ 
\hline

$\frac{\lambda v^2}{2-m^2}$ & $ \frac{2}{2-m^2}$   & $ m^2-1$ & $ \text{dn}\left(w+d,m\right)$  & 
$\pm\,v\,\varepsilon\, \text{dn}\left(w+d,m\right)$ 
& $ {1-m^2}$    \\ 
\hline

$-\frac{\lambda v^2}{1+m^2}$ &  $\frac{2m^2}{1+m^2}$   & $1$ & $ \text{cd}\left(w,m\right)$ 
& $\pm\,v\,\varepsilon\, \text{cd}\left(w+d,m\right)$  
&  $m^2$ \\ 
\hline

$\frac{\lambda v^2}{2m^2-1}$ &  $\frac{2m^2(1-m^2)}{2m^2-1}$   & $1$ & $ \text{sd}\left(w,m\right)$ 
& $\pm\,v\,\varepsilon\, \text{sd}\left(w+d,m\right)$ 
&  $\frac{m^2-1}{m^2}$  \\ 
\hline

$\frac{\lambda v^2}{2-m^2}$ &  $\frac{2(1-m^2)}{2-m^2}$   & $-1$ & $ \text{nd}\left(w,m\right)$ 
& $\pm\,v\,\varepsilon\, \text{nd}\left(w+d,m\right)$  
 & $ {1-m^2}$   \\ 
\hline

$-\frac{\lambda v^2}{1+m^2}$ &  $\frac{2}{1+m^2}$   & $m^2$ & $ \text{dc}\left(w,m\right)$ 
& $\pm\,v\,\varepsilon\, \text{dc}\left(w+d,m\right)$ 
 & $m^2$  \\ 
\hline

$\frac{\lambda v^2}{2m^2-1}$ &  $-\frac{2(1-m^2)}{2m^2-1}$   & $-m^2$ & $ \text{nc}\left(w,m\right)$ 
& $\pm\,v\,\varepsilon\, \text{nc}\left(w+d,m\right)$  
&  $\frac{m^2-1}{m^2}$   \\ 
\hline

$\frac{\lambda v^2}{2-m^2}$ &  $-\frac{2(1-m^2)}{2-m^2}$   & $1$ & $ \text{sc}\left(w,m\right)$ 
& $\pm\,v\,\varepsilon\, \text{sc}\left(w+d,m\right)$ 
 &   $ {1-m^2}$   \\ 
\hline

$-\frac{\lambda v^2}{1+m^2}$ &  $\frac{2}{1+m^2}$   & $m^2$ & $ \text{ns}\left(w,m\right)$ 
& $\pm\,v\,\varepsilon\, \text{ns}\left(w+d,m\right)$  
 & $m^2$   \\ 
\hline

$\frac{\lambda v^2}{2m^2-1}$ &  $\frac{-2}{2m^2-1}$   & $-m^2(1-m^2)$ & $ \text{ds}\left(w,m\right)$ 
& $\pm\,v\,\varepsilon\, \text{ds}\left(w+d,m\right)$ 
 &  $\frac{m^2-1}{m^2}$  \\ 
\hline

$\frac{\lambda v^2}{2-m^2}$ &  $\frac{-2}{2-m^2}$   & $1-m^2$ & $ \text{cs}\left(w,m\right)$ 
& $\pm\,v\,\varepsilon\, \text{cs}\left(w+d,m\right)$  
 &  $ {1-m^2}$   \\ 
\hline

\end{tabular}
\caption{The twelve Jacobi elliptic functions solutions to  (\ref{dy2-1}) and  (\ref{phi-four-2})  and their corresponding parameters 
$p^2$ (the mass-shell relation),  $\varepsilon^2$ and $\frac{a}{\varepsilon^2}$. The table gives also the relation between $k^2$ and 
$m^2$ appearing in (\ref{sol-con}).}
\label{tab1-1}
\end{table}
\end{center}
It is important to notice that  $\varepsilon^2$ has to be strictly positive
(for $\rho(w)$ to be a real field and different from zero). This, consequently, puts 
restrictions on the allowed values of the parameter $m$ for some of the solutions.

\section{Solutions to \texorpdfstring{$\phi^4$}{phi} in terms of trigonometric and hyperbolic functions}

We have seen that the equation of motion for the field $\rho(w)$ is given by the 
differential equation
\bea
\frac{\text{d}^2\rho}{\text{d}w^2} + \frac{\lambda}{p^2}\,\rho\left(\rho^2 -v^2\right)
 &=& 0\,\,\,\,\,\,.
\eea
The general solution to this equation is expressed in terms of the twelve Jacobi elliptic functions depending 
on a  parameter $m$. On the other hand, the Jacobi elliptic functions reduce to ordinary
trigonometric or hyperbolic functions for the two special values $m=0$ and $m=1$ \cite{gradshteyn,abramowitz,nist}.

\par
There are twelve different trigonometric and  hyperbolic functions corresponding to the values  $m=0$ and $m=1$.
However, only five\footnote{The other seven functions, for $m=0$ and $m=1$, lead either to 
$\rho(w)=0$ or to a complex $\rho(w)$. {}For example, 
${\text{ds}}{(w+d\,,\,1)}= \frac{1}{\sinh{(w+d)}}$ but $\varepsilon^2=-2$ for $m=1$, as can be seen 
from table \ref{tab1-1}.
This leads to a complex scalar field $\rho(w)$.} of them are solutions 
the above differential equation. These are
\bea
\rho\left(w\right) &=& \pm v\,{\text {sn}}\left[\tau\left(w+\alpha\right)+\beta\,,1\,\right]
=\pm v\, \tanh\left[\tau\left(w+\alpha\right)+\beta\right]
\,\,\,\,\,\,\,\,\,,\,\,\,\,\,\, p^2=-\frac{\lambda\,v^2}{2\tau^2}\,\,,
\label{hyper-1}
\\
\rho\left(w\right) &=& \pm \,v\,{\text {ns}}\left[\tau\left(w+\alpha\right)+\beta\,,1\,\right]
=\frac{\pm \,v\,}{\tanh \left[\tau\left(w+\alpha\right)+\beta\right]}
\,\,\,\,\,\,\,\,\,,\,\,\,\,\,\,\,\,\, p^2=-\frac{\lambda\,v^2}{2\tau^2}\,\,\,\,,
\\
\rho\left(w\right) &=& \pm \sqrt{2}\,v\,{\text {cn}}\left[\tau\left(w+\alpha\right)+\beta\,,1\,\right]
=\frac{ \pm \sqrt{2}\,v\,}{\cosh \left[\tau\left(w+\alpha\right)+\beta\right]}
\,\,\,\,\,\,\,\,\,,\,\,\,\,\,\,\,\,\, p^2=\frac{\lambda\,v^2}{\tau^2}\,\,\,\,,
\\
\rho\left(w\right) &=& \pm \sqrt{2}\,v\,{\text {ds}}\left[\tau\left(w+\alpha\right)+\beta\,,0\,\right]
=\frac{ \pm \sqrt{2}\,v\,}{\sin \left[\tau\left(w+\alpha\right)+\beta\right]}
\,\,\,\,\,\,\,\,\,,\,\,\,\,\,\,\,\,\, p^2=-\frac{\lambda\,v^2}{\tau^2}\,\,\,\,,
\\
\rho\left(w\right) &=& \pm \sqrt{2}\,v\,{\text {dc}}\left[\tau\left(w+\alpha\right)+\beta\,,0\,\right]
=\frac{ \pm \sqrt{2}\,v\,}{\cos \left[\tau\left(w+\alpha\right)+\beta\right]}
\,\,\,\,\,\,\,\,\,,\,\,\,\,\,\,\,\,\, p^2=-\frac{\lambda\,v^2}{\tau^2}\,\,.
\eea
The solutions depend on a parameter $\tau$ and a constant $\tau\alpha +\beta$. However, 
sometimes these solutions are found under different writings \cite{bekir,demiray-bulut,wazwaz,esen}. {}For instance, the first
solution  (\ref{hyper-1}) can be expressed as
\bea
\rho\left(w\right) &=& \pm v\, \tanh\left[\tau\left(w+\alpha\right)+\beta\right]
= \pm v\,\frac{\mu+ \tanh\left[\tau\left(w+\alpha\right)\right] }
{1+ \mu\tanh\left[\tau\left(w+\alpha\right)\right] }
\,\,,\,\,\mu=\tanh\left(\beta\right)\,\,.
\eea
Other expressions can be reached  by simply expanding the arguments of the trigonometric 
and hyperbolic functions.


\begin{thebibliography}{99}


\bibitem{nielsen-olesen}
H. B. Nielsen and P. Olesen, {\it Vortex-line models for dual strings},  Nucl. Phys. B {\bf 61} (1973)
45-61.

\bibitem{witten}
Edward Witten,  {\it Superconducting strings}, Nucl. Phys. B {\bf 249}, Issue 4,  (1985) 557-592.




\bibitem{bogomolny}
E. B. Bogomolny, {\it The stability of classical solutions}, Soviet
Journal of Nuclear Physics, vol. {\bf 24},  (1976) pp. 449–454.



\bibitem{vilenkin-shellard}
A. Vilenkin and  E. P. S. Shellard, {\it Cosmic strings and other topological defects}, in Cambridge
Monographs in Mathematical Physics, Cambridge University Press (2000).


\bibitem{manton-sutcliffe}
N. Manton and P. Sutcliffe, {\it Topological solitons}, Cambridge University Press (2004).

\bibitem{klacka-et-al}
J. Kla\v{c}ka, Metod Saniga and J\'{a}n Ryb\'{a}k, 
{\it Numerical analysis of a static cylindrically symmetric Abelian Higgs sunspot},
Contributions of the Astronomical Observatory Skalnate Pleso, vol. 22, p. 107-115 (1992).


\bibitem{lee-pang}
T. D. Lee and Y. Pang, {\it Nontopological solitons},  Phys. Rept. {\bf 221}, (1992) 251-350.

\bibitem{shnir}
Ya. M. Shnir, {\it Topological and non-topological solitons in scalar field theories},  Cambridge
University Press (2018).

\bibitem{nugaev-et-al}
 E. Ya. Nugaev and A. V. Shkerin, {\it Review of nontopological solitons in theories with $U(1)$
symmetry}, JETP {\bf 130}, (2020) 301-320, arXiv:1905.05146 [hep-th].


\bibitem{volkov-radu}
Eugen Radu and Mikhail S. Volkov, {\it  Stationary ring solitons in field theory-knots and vortons}, 
Phys.  Rept. {\bf 468} (2008) 101–151, arXiv:0804.1357 [hep-th].


\bibitem{brihaye-et-al}
Yves Brihaye, Stefan Giller, Piotr Kosinski and  Jutta Kunz,
{\it Sphalerons and normal modes in the (1+1)-dimensional Abelian Higgs model on the circle}, 
Phys. Lett. {\bf B 293} (1992) 383-388.


\bibitem{manton-samols}
N. S. Manton and T. M. Samols, {\it Sphalerons on a circle}, 
Phys. Lett. {\bf B 207} (1988) 179.

\bibitem{canfora-1}
F. Canfora, A. Cisterna, D.  Hidalgo and J. Oliva,
{\it Exact $pp$-waves, (A)dS waves and Kundt spaces in the
Abelian-Higgs model}, Phys .Rev. {\bf D 103} (2021) 8, 085007, arXiv:2102.05481 [hep-th].
\bibitem{canfora-2}
F.  Canfora, 
{\it Ordered arrays of Baryonic tubes in the Skyrme model in $( 3+1)$ dimensions at finite density},
Eur. Phys. J. {\bf C 78} (2018) 11, 929, 1807.02090 [hep-th].
\bibitem{canfora-3}
L. Avil\'es, F. Canfora, N. Dimakis, and D. Hidalgo, 
{\it Analytic topologically nontrivial solutions of the $(3+1)$-dimensional $U(1)\times U(1)$ gauged Skyrme model and extended duality}, 
Phys. Rev. {\bf D 96} (2017) 12, 125005, 1711.07408 [hep-th].
\bibitem{canfora-4}
F. Canfora, M. Lagos, S. H. Oh, J. Oliva and A. Vera, 
{\it Analytic $(3+1)$-dimensional gauged Skyrmions, Heun, and Whittaker-Hill equations and resurgence},
Phys. Rev. {\bf D 98} (2018) 8, 085003, 1809.10386 [hep-th].
\bibitem{canfora-5}
F. Canfora, N. Dimakis, and A. Paliathanasis, 
{\it  Analytic Studies of Static and Transport Properties of (Gauged) Skyrmions},
Eur. Phys. J. {\bf C 79} (2019) 2, 139,  1902.01563 [hep-th].
\bibitem{canfora-6}
F. Canfora, S. H. Oh and A. Vera, 
{\it Analytic crystals of solitons in the four dimensional gauged non-linear sigma model},
 Eur. Phys. J. {\bf C 79} (2019) 6, 485, 1905.12818 [hep-th].
\bibitem{canfora-7}
F. Canfora, M. Lagos and A. Vera,
{\it Crystals of superconducting Baryonic tubes in the low energy limit of QCD at finite density}, 
Eur. Phys. J. {\bf C 80} (2020) 8, 697,  2007.11543 [hep-th].
\bibitem{canfora-8}
F. Canfora, A. Giacomini, M. Lagos, S. H. Oh, A. Vera, 
{\it Gravitating superconducting solitons in the $(3+1)$-dimensional Einstein gauged non-linear $\sigma$-model},
Eur. Phys. J. {\bf C 81} (2021) 1, 55, 2001.11910 [hep-th].


\bibitem{brihaye}
Y. Brihaye, 
{\it Non-Abelian Plane Waves in the Higgs Model}, Lett. Nuovo. Cim. {\bf 36} (1983) 275.



\bibitem{diakonos-et-al}
F. K. Diakonos, G. C. Katsimiga, X. N. Maintas and C. E. Tsagkarakis,
{\it Symmetric solitonic excitations of the $(1 + 1)$-dimensional Abelian-Higgs classical vacuum},
Phys. Rev. {\bf  E 91} (2015) 2, 023202, 1404.1607 [hep-th].



\bibitem{achellios-et-al}
V. Achilleos, F. K. Diakonos, D. J. Frantzeskakis, G. C. Katsimiga, X .N. Maintas, E. Manousakis, C. E. Tsagkarakis and A. Tsapalis,
{\it Oscillons and oscillating kinks in the Abelian-Higgs model}, 
Phys. Rev. {\bf D 88}, 045015 (2013), arXiv:1306.3868 [hep-th].


\bibitem{katsimiga}
G. C. Katsimiga, F. K. Diakonos and X. N. Maintas,
{\it Classical dynamics of the Abelian Higgs model from the critical point and beyond},
Phys. Lett. {\bf B 748} (2015) 117-124.





\bibitem{rozowsky-et-al}
J. S. Rozowsky, R. R. Volkas and  K. C. Wali, {\it Domain wall solutions with Abelian gauge fields}, 
Phys. Lett. {\bf B 580} (2004) 249-256,  arXiv:hep-th/0305232.

\bibitem{george-volkas}
Damien P. George and Raymond R. Volkas, {\it Stability of domain walls coupled to Abelian gauge fields},
Phys. Rev. {\bf D 72} (2005) 105011, 	arXiv:hep-ph/0508206.


\bibitem{bazeia}
D. Bazeia, L. Losano, M. A. Marques and  R. Menezes,
{\it Analytic vortex solutions in generalized models of the Maxwell-Higgs type}, 
Phys. Lett. {\bf B 778} (2018) 22, arXiv:1801.01077 [hep-th].

\bibitem{casana}
R. Casana, M. M. Ferreira Jr., E. da Hora and C. dos Santos,
{\it Analytical BPS Maxwell-Higgs vortices}, Advances in High Energy Physics, {\bf vol.  2014}, Article ID 210929, (2014), 
arXiv:1405.7920 [hep-th].







\bibitem{gradshteyn}
{\it Table of Integrals, Series, and Products}, I. S. Gradshteyn and I. M. Ryzhik,
(Alan Jeffrey, Editor), Academic Press, Fifth Edition (1994). 


\bibitem{abramowitz}
{\it Handbook of Mathematical Functions}, Edited by  M. Abramowitz and I. A. Stegun,
Dover Publications, New York, Ninth Printing (1970).


\bibitem{nist}
{\it NIST Handbook of Mathematical Functions}, Edited by F. W. J. Olver, D. W. Lozier, R. F. Boisvert and C. W. Clark,
Cambridge University Press,  New York, First Published (2010).




\bibitem{valent}

Galliano Valent,
{\it Heun functions versus elliptic functions},  arXiv:math-ph/0512006v1, (2005). 


\bibitem{pastras}

Georgios Pastras, {\it Four Lectures on Weierstrass Elliptic Function and
Applications in Classical and Quantum Mechanics},  arXiv:math-ph/1706.07371v2, (2017).



\bibitem{khaled}

Khaled A. Gepreel, 
{\it Explicit Jacobi elliptic exact solutions for nonlinear partial fractional differential equations}, 
Advances in Difference Equations, Vol. 2014, 286-300, (2014).


\bibitem{turks}

A. Da\c{s}cio\u{g}lu, S. \c{C}ulha \"{U}nal and D. Varol Bayram, 
{\it New Analytical Solutions for Space and Time Fractional Phi-4 Equation}, 
NATURENGS, MTU Journal of Engineering and Natural Sciences 1:1 (2020) 30-46.





\bibitem{frasca}

Marco Frasca,
{\it Exact solutions of classical scalar field equations},  Journal of Nonlinear Mathematical Physics, Vol. 1, No. 1 (2009) 1–7.




\bibitem{bekir}
A. Bekir, {\it New Exact Travelling Wave
Solutions for Regularized Long-wave, Phi-Four
and Drinfeld-Sokolov Equations}, International
Journal of Nonlinear Science, 6(1),  (2008) 46-52.



\bibitem{demiray-bulut}
Seyma Tuluce Demiray, Hasan Bulut, {\it Analytical solutions of Phi-four equation },
An International Journal of Optimization and Control: Theories \& Applications, 
Vol.7, No.3, (2017) 275-280. 




\bibitem{wazwaz}
A. M. Wazwaz, {\it A sine-cosine method for handling nonlinear wave equations},
Mathematical and Computer Modelling, 40, (2004) 499-
508.


\bibitem {esen}
Berat Karaagac, Selcuk Kutluay, Nuri Murat Yagmurlu and Alaattin Esen,
{\it Exact solutions of nonlinear evolution equations using the
extended modified Exp(-$\Omega(\xi)$) function method},
Tbilisi Mathematical Journal 12(3), (2019) 109–119.




\bibitem {liang-et-al}
Jiu-Qing Liang, H. J. W. M\"{u}ller-Kirsten and D. H. Tchrakian, 
{\it Solitons, bounces and sphalerons on a circle}, Phys. Lett. {\bf B 282 } (1992) 105-110.


\end{thebibliography}
\end{document}